\documentclass[10pt]{article}

\usepackage{graphicx}
\usepackage{amsmath}
\usepackage[title]{appendix}
\usepackage[top=1in, bottom=1.25in, left=1.00in, right=1.00in]{geometry}
\usepackage{tabularx}
\usepackage{hyperref}

\title{A new and stable estimation method of country economic fitness and product complexity}

\author{
Vito~D.~P.~Servedio $^{1}$,
Paolo Butt\`a $^{2}$,
Dario Mazzilli $^{3}$,\\
Andrea Tacchella $^{4}$,
Luciano Pietronero  $^{3}$\\[5mm]
{ \small
	$^{1}$ Complexity Science Hub Vienna, Josefst\"atter-Strasse 39, A-1080 Vienna, Austria}\\
	{\small
	$^{2}$ Department of Mathematics, Sapienza University of Rome, P.le Aldo Moro 5, 00185 Roma, Italy}\\
	{\small
	$^{3}$ Physics Department, Sapienza University of Rome, P.le Aldo Moro 5, 00185 Roma, Italy}\\
	{\small
	$^{4}$ Institute for Complex Systems, CNR, Via dei Taurini 19, Rome, Italy}\\
{\small Correspondence: v.servedio@gmail.com}
}

\begin{document}
\maketitle
\abstract{
\noindent
We present a new metric estimating fitness of countries and complexity of products by exploiting a non-linear non-homogeneous map applied to the publicly available information on the goods exported by a country.
The non homogeneous terms guarantee both convergence and stability.
After a suitable rescaling of the relevant quantities, the non homogeneous terms are eventually set to zero so that this new metric is parameter free.
This new map almost reproduces the results of the original homogeneous metrics already defined in literature and allows for an approximate analytic solution in case of actual binarized matrices based on the Revealed Comparative Advantage  (RCA) indicator.
This solution is connected with a new quantity describing the neighborhood of nodes in bipartite graphs, representing in this work the relations between countries and exported products.
Moreover, we define the new indicator of country \emph{net-efficiency} quantifying how a country efficiently invests in capabilities able to generate innovative complex high quality products.
Eventually, we demonstrate analytically the local convergence of the algorithm involved.\\[3mm]
}

\section{Introduction}

In the last decade a new approach to macroeconomics has been developed to better understand the growth of countries~\cite{tacchella2012, hidalgo2009}.
The key idea is to consider the international trade of countries as a proxy of their internal production system.
By describing the international trade as a  bipartite network, where countries and products are sites of the two layers, new metrics for the economy of countries and the quality of products can be constructed by leveraging the network structures only~\cite{tacchella2012} .
These metrics quantify the fitness of countries, the quality of their industrial system, and the complexity of commodities by indirectly inferring the technological requirements needed to produce them. The mathematical properties of the algorithm involved in the evaluation of the metrics, as well as the economic meaning of the metrics and possible applications, have been discussed in previous papers~\cite{tacchella2013economic, cristelli2013measuring, zaccaria2016case}. Moreover, these two new metrics have been successfully used to develop state-of-the-art forecasting approaches for economic growth~\cite{Saracco2016,cristelli2017predictability,tacchella2018dynamical}.

The very same approach has been applied to different social and ecological systems presenting a bipartite network structure and a competition between the components of the system~\cite{munoz2015,gabrielli2014}.
Thus, it is natural to interpret fitness and complexity as properties of the network underlying those systems.
The revised version of the fitness-complexity estimation metric that we show here, results in a clear and natural interpretation in terms of network properties and helps to better understand the different components that contribute to the fitness of countries.

In the following, we first describe the original metrics of fitness and complexity and their properties, underlining some critical issues that we solve with this new version.
Then, we define the new procedure step by step and highlight its advantages in the case of countries-products bipartite networks.
Finally, we devise an approximated solution and discuss its interpretation.
In Appendix~\ref{app:table} we list the main quantities appearing in the text.

\section{Metric Definition}

\subsection{The Original Metric}

Object of this work is the network of countries and their exported goods.
This network is of bipartite type (countries and products are mutually linked, but no link exists between countries as well as between products) and weighted (links carry a weight $s_{cp}$, i.e., the exported volume of product $p$ of country $c$, measured in US\$).
Data ranging from year 1995 to year 2015 can be freely retrieved from the Web \cite{comtrade}, though we use them after a procedure to enhance their consistency \cite{tacchella2018dynamical}.
Eventually, we come up with data about 161 countries and more than 4000 products, which were categorized according to the Harmonized System 2007 coding system, at 6 digits level of coarse-graining.
The weighted bipartite network of countries and products can be projected onto an unweighted network described solely by the $M_{cp}$ matrix  with elements set to unity when a given country $c$ meaningfully exports a good $p$ and zero otherwise (See Methods).

The original metric estimating the fitness of countries and complexity of products was defined by the following non-linear iterative map:
\begin{equation}
\displaystyle
  \left\{
    \begin{array}{ll}
      F_c^{(n)} = \sum_{p'} M_{cp'} Q_{p'}^{(n-1)} & \mbox{with}~~~ 1\le c \le \mathcal{C}\\
      Q_p^{(n)} = \left(\sum_{c'} M_{c'p}/F_{c'}^{(n-1)}\right)^{-1} & \mbox{with}~~~ 1\le p \le \mathcal{P},\\
    \end{array}
  \right.
  \label{eq:oldalg}
\end{equation}
with initial values $F_c^{(0)}=Q_p^{(0)}=1, \forall\, c, p$.
In the previous expression, $F_c$ and $Q_p$ stand for the fitness of a country $c$ and quality (complexity) of a product $p$; $\mathcal{C}$ and $\mathcal{P}$ are the total number of countries and exported products, respectively; and from the dataset we have that $\mathcal{C}\ll\mathcal{P}$.

By multiplying all $F_c$ and $Q_p$ by the same numerical factor $k$, the map remains unaltered, so that the fixed point of the map (as $n\rightarrow\infty$) is defined up to a normalization constant.
In the original method this constant is chosen at each iteration $n$ such that fitness and complexity are constrained to lie on the double simplex defined by:
\begin{equation}
 \sum_c F_c^{(n)} = \mathcal{C} ~~~\mbox{and}~~~ \sum_p Q_p^{(n)} = \mathcal{P}.
  \label{eq:normalization}
\end{equation}
The metric defined in Equations~(\ref{eq:oldalg}) and (\ref{eq:normalization}) successfully ranks the countries of our world according to their potential technological development and, when applied to different yearly time intervals, can be used to suggest precise strategies to improve country economies.
It has also been proved to give the correct ranking of importance of species in a complex ecological system \cite{munoz2015}.
Despite its success, some points can  still be improved:
\begin{itemize}
 \item[i.] {Convergence issues}: 
    As stated in a recent paper dealing with the stability of calculating this metric~\cite{Pugliese2016}:
    \begin{quote}
      If the belly of the matrix [$M_{cp}$] is outward, all the fitnesses and complexities converge to numbers greater than zero. If the belly is inward, some of the fitnesses will converge to zero.
    \end{quote}
    Since an inward belly is the rule rather than the exception, some countries will have zero fitness and as a result all the products exported by them get zero complexity (quality).
    This is mathematically acceptable, though it heavily underestimates the quality of such products: Even natural resources need the right know-how to be extracted so that their quality would be better represented by a positive quantity. To cure this issue one has to introduce the notion of ``rank convergence'' rather than absolute convergence, i.e., the fixed point is considered achieved when the ranking of countries stays unaltered step by step.
 \item[ii.] {Zero exports}:
    The countries that do not export any good do have zero fitness independently from their finite capabilities.
 \item[iii.] {Specialized world}:
    In an hypothetical world where each country would export only one product, different from all other products exported by other countries, this metric would assign a unity fitness and quality to all countries and products. Though mathematically acceptable, this solution does not take into account the intrinsic complexity of products.
 \item[iv.] {Equation symmetry}:
    This is rather an aesthetic point, in that Equation~(\ref{eq:oldalg}) are not cast in a symmetric form.
\end{itemize}

\subsection{The New Metric}

First, we reshape Equation~(\ref{eq:oldalg}) in a symmetric form by introducing the variable $P_p=Q_p^{-1}$, i.e.:
\begin{equation}
 \displaystyle
  \left\{
    \begin{array}{ll}
      F_c^{(n)} = \sum_{p'} M_{cp'}/P_{p'}^{(n-1)} & \mbox{with}~~~ 1\le c \le \mathcal{C}\\
      P_p^{(n)} = \sum_{c'} M_{c'p}/F_{c'}^{(n-1)} & \mbox{with}~~~ 1\le p \le \mathcal{P}.\\
    \end{array}
  \right.
\end{equation}
Now the quality of products are given by the quantities $P_p^{-1}$ and the metric is trivially equivalent to the original one provided one uses the normalization conditions $\sum_c F_c^{(n)} = \mathcal{C}$ and $\sum_p (P_p^{(n)})^{-1} = \mathcal{P}$.

Next, we introduce two set of quantities $\phi_c>0$ and $\pi_p>0$ and consider the inhomogeneous non-linear map defined as:
\begin{equation}
  \displaystyle
  \left\{
    \begin{array}{ll}
      F_c^{(n)} = \phi_c+ \sum_{p'} M_{cp'}/P_{p'}^{(n-1)} & \mbox{with}~~~ 1\le c \le \mathcal{C}\\
      P_p^{(n)} = \pi_p+ \sum_{c'} M_{c'p}/F_{c'}^{(n-1)} & \mbox{with}~~~ 1\le p \le \mathcal{P}.\\
    \end{array}
  \right.
\label{eq:newalg}
\end{equation}
Since the map is no more defined up to a multiplicative constant, the normalization condition is not required anymore, while the initial condition can be set as in the original metric $F_c^{(0)}=P_p^{(0)}=1, \forall\, c, p$.
The fixed point of the transformation is now trivially characterized by the conditions:
\begin{equation}
 F_c \ge \phi_c, ~~~ P_p \ge \pi_p, ~~~ F_c P_p > M_{cp}.
\end{equation}
The parameters $\phi_c$ and $\pi_p$ can be interpreted as follows.
The parameter $\phi_c$ represents the intrinsic fitness of a country.
In fact, for a country $k$ that does not export any good we have $M_{kp}=0~\forall p$ so that its fitness is simply equal to $\phi_k$. Irrespective of its exports any country has a set of capabilities that characterize it.

The parameter $\pi_p$ is more intriguing.
If no country exports it (probably because no country produces it), the product $q$ has not been invented yet and its quality lies at its maximum value $\pi_q^{-1}$ since $M_{cq}=0~\forall c$. Therefore, the inverse of $\pi_q$ may be interpreted as a sort of innovation threshold: The smaller the parameter is, the higher is the quality of the product in his outset and more sophisticated capabilities are necessary to produce it.
On the other hand, products like natural resources may be associated with a larger value of the parameter since require less complex capabilities for their extraction.

In order to keep the metric evaluating algorithm simple and parameter free as in the original case, we set a common value $\phi_c=\pi_p=\delta$, then we study the dependence of the metrics on $\delta$, and finally we set $\delta=0$ (in fact renouncing to cure the issue number iii.\ listed above).

\section{Results}
\subsection{Dependence on the Non-Homogeneous Parameter}

\begin{figure}[t]
\centering
 \includegraphics[width=8cm]{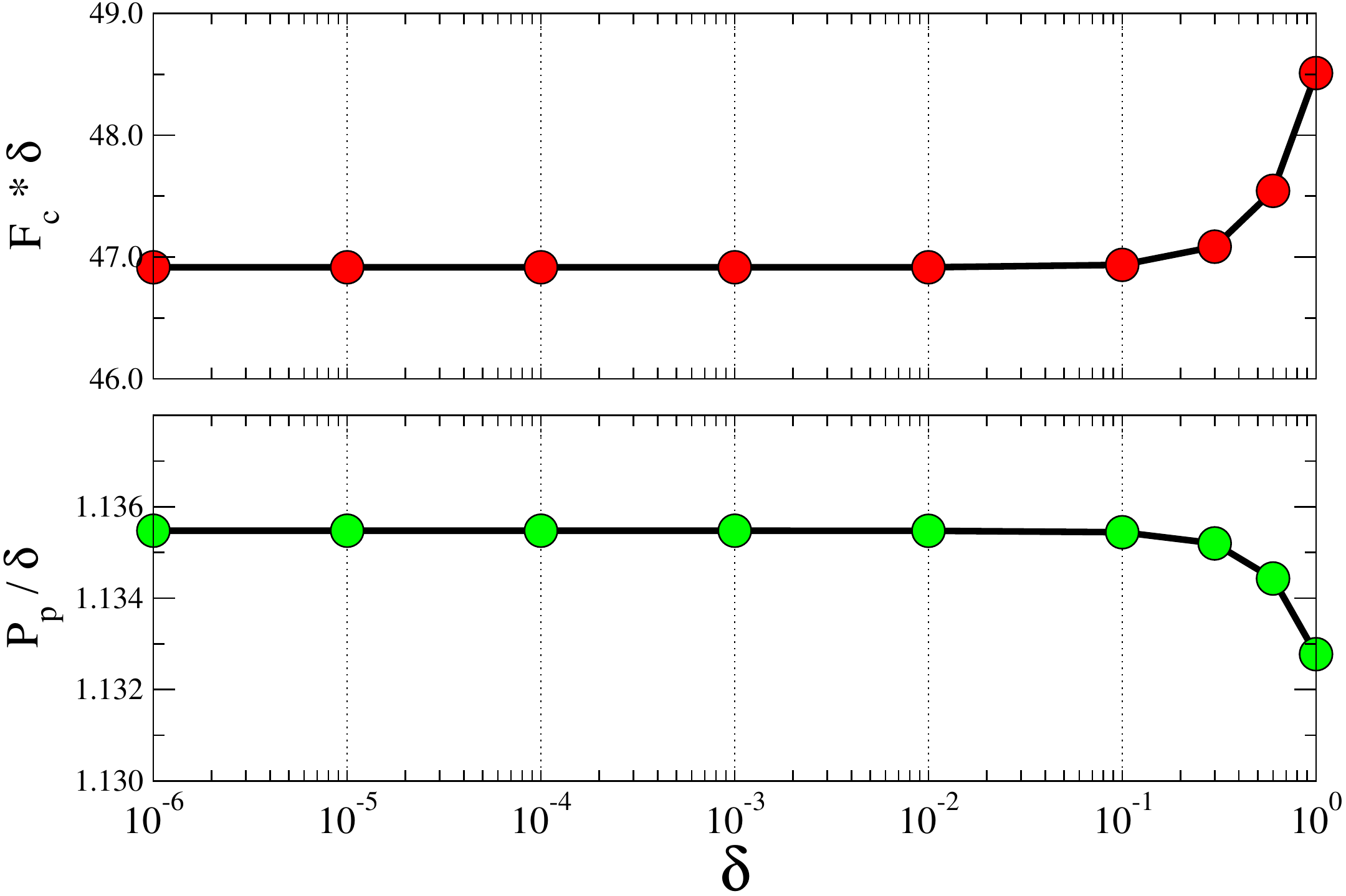}
\caption{ {Dependence on the non-homogeneous parameter:}
  Dependence of fitness and quality at the fixed point on the parameter $\delta$. One country (Afghanistan) and one product (live horses) were chosen arbitrarily from the sample of year 2014.
  \label{fig:dep_on_delta}
}
\end{figure}

\begin{figure}[t]
\centering
 \includegraphics[width=0.49\textwidth]{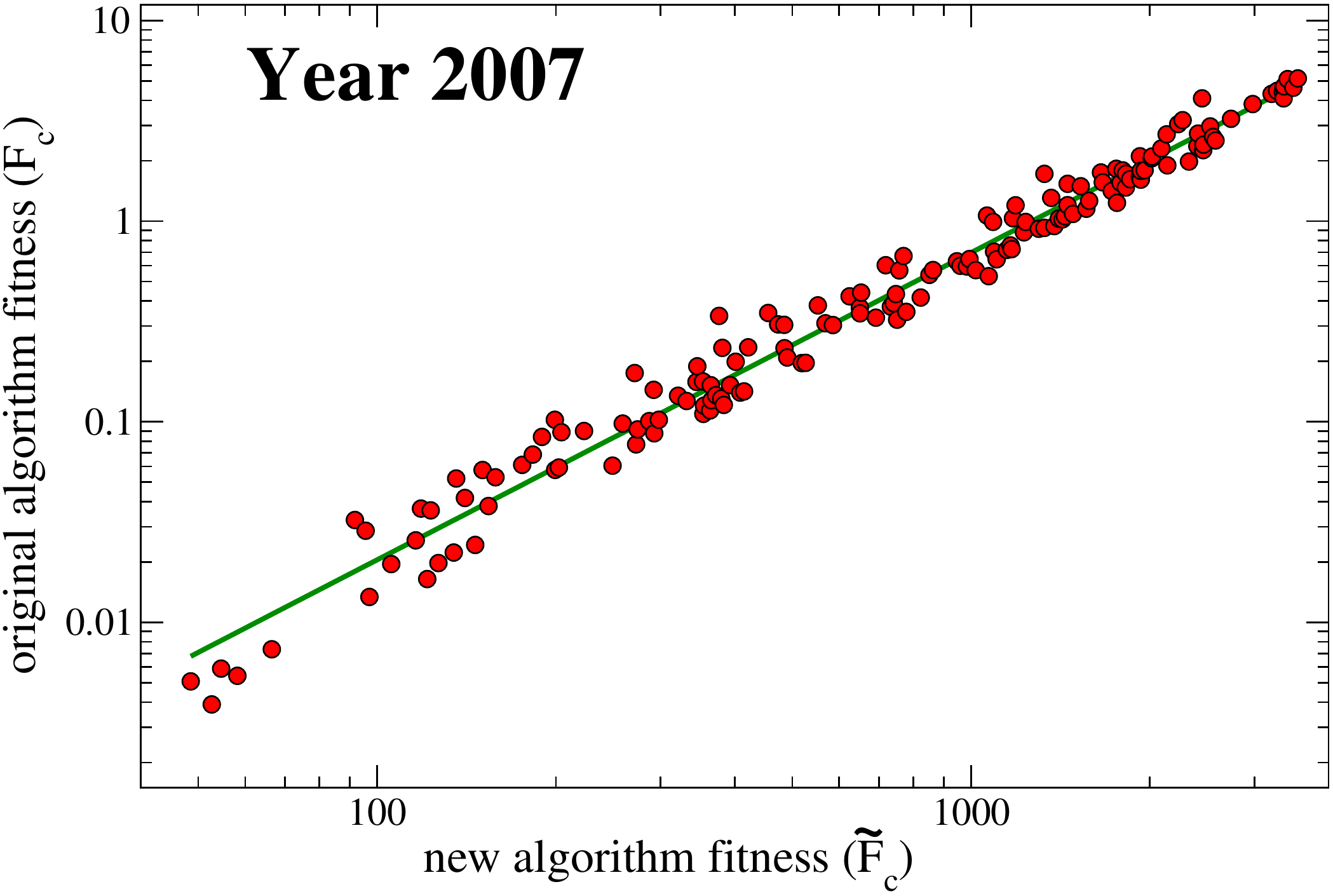}%
 \includegraphics[width=0.49\textwidth]{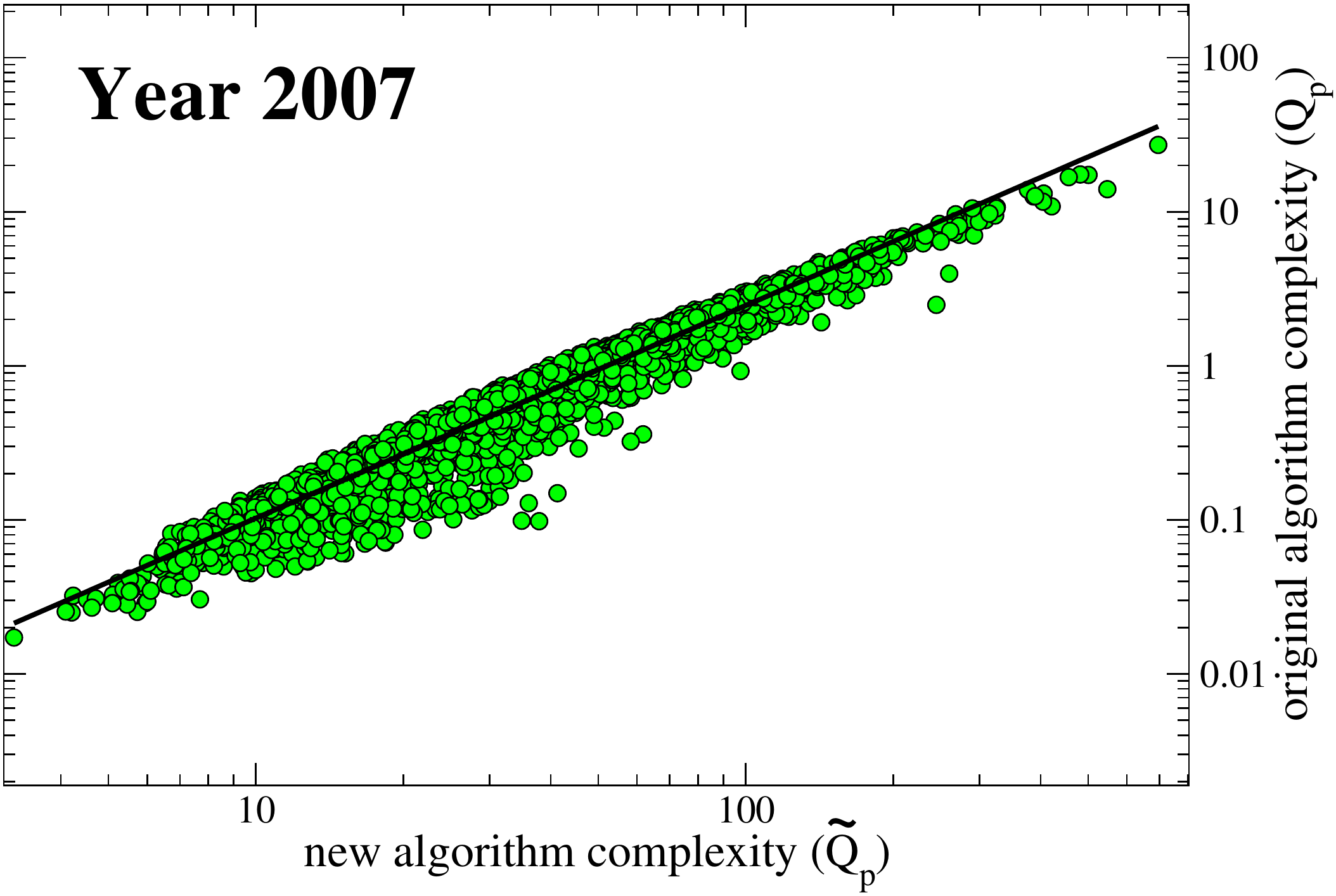}
\caption{ {Comparison between the original and the revised metric:}
  Differences in country fitness (left panel) and product complexity (right panel) calculated in the original metric of Ref.~\cite{tacchella2012} (vertical axes) and new metric (horizontal axes) as referred to year 2007.
  The green line in the left panel is the best least square approximation of power-law type (correlation coefficient 0.989) with exponent ca.~1.53. The dark line in the right panel is the best power-law approximation (correlation coefficient 0.971) resulting with an exponent of ca.~1.38.
  The year 2007 was chosen randomly. Similar results apply to all the years considered.
  In particular, the correlation coefficient and the exponent of the green line in the left panel lie between 0.987 and 0.990, and 1.48 and 1.61 respectively throughout the years.
  For the black line in the right panel we find a correlation coefficient between 0.950 and 0.979, and an exponent between 1.34 and 1.47.
  \label{fig:comparison}
}
\end{figure}

We consider $\phi_c=\pi_p=\delta\ \forall c,p$ and address the dependence of the fixed point upon $\delta$.
To outline the dependence of $F_c$ and $P_p$ from the parameter $\delta$, we  use the relations defined in Equation~(\ref{eq:newalg}) and introduce the rescaled quantities $\tilde{P}_p = P_p/\delta$ and $\tilde{F}_c= F_c \delta$.
After some trivial algebra we get from Equation~(\ref{eq:newalg}):
\begin{equation}
  \displaystyle
  \left\{
    \begin{array}{ll}
      \tilde{F}_c^{(n)} = \delta^2+ \sum_{p'} M_{cp'}/\tilde{P}_{p'}^{(n-1)} & \mbox{with}~~~ 1\le c \le \mathcal{C}\\
      \tilde{P}_p^{(n)} = 1+ \sum_{c'} M_{c'p}/\tilde{F}_{c'}^{(n-1)} & \mbox{with}~~~ 1\le p \le \mathcal{P},\\
    \end{array}
  \right.
\label{eq:newalg2}
\end{equation}
from which we deduce that, as soon as the parameter $\delta^2$ is much smaller than the typical value of $M_{cp}$ matrix elements, i.e., much smaller than unity, the fixed point in terms of $\tilde{F}_c$ and $\tilde{P}_p$ almost does not depend on $\delta$ (see Figure~\ref{fig:dep_on_delta}).
It is worth noting that the values of fitness $F_c$ and quality $Q_p = P_p^{-1}$ of the original map defined by Equations~(\ref{eq:oldalg}) and (\ref{eq:normalization}) cannot be obtained in this new metric when the parameter $\delta$ tends to zero.
In terms of $\tilde{F}_c$ and $\tilde{P}_p$ the fitness and quality obtained in the original metric can be expressed as $F_c = \tilde{F}_c\,\delta^{-1}$ and $Q_p = \tilde{P}_p^{-1} \,\delta^{-1}$.
Since the new metric provides finite non vanishing values of $\tilde{F}_c$ and $\tilde{P}_p$, by taking the limit $\delta\rightarrow 0$ would deliver infinite values of $F_c$ and $Q_p$.
We might think that the normalization procedure necessary in the old metric in order to fix the arbitrary constant would get rid of the common factor $\delta^{-1}$ and deliver the same values of the new metric.
Unfortunately, this is not the case since the new metric does not rely on a normalization procedure.
Therefore, since a self-consistent procedure of normalization, i.e., a projection on the double simplex defined by Equation~(\ref{eq:normalization}), is missing in the new metric, the results cannot coincide.
Since the quantities $\tilde{F}_c$ and $\tilde{P}_p$ are well defined in the limit $\delta\rightarrow0$, we shall focus on them only, in the following.
We remind that the complexities of products delivered by the original metric are connected to the set of $P_p^{-1}$ and thus to the $\tilde{P}_p^{-1}$.
In particular, the second of Equation~(\ref{eq:newalg2}) can be interpreted at the fixed point as:
\(
\tilde{P}_p = 1 + \tilde{Q}_p^{-1}
\)
with the $\tilde{Q}_p$ expressed as in the second of Equation~(\ref{eq:oldalg}), but with the tilde quantities calculated in the new metric.
Therefore, we shall assign to $\tilde{Q}_p = (\tilde{P}_p-1)^{-1}$ the meaning of complexity of products in our new metric.
The differences between the old and new metrics are depicted in Figure~\ref{fig:comparison}.

\subsection{Analytic Approximate Solution}

In this section we shall provide an approximate analytic solution that can be used to estimate the values attained by the map of Equation~(\ref{eq:newalg2}) at the fixed point.
Despite their symmetric shape, Equation~(\ref{eq:newalg}) are not symmetric at all since in case of actual countries and products, the matrix $M_{cp}$ is rectangular with the number of its rows $\cal{C}$ being much less than the number of its columns $\cal P$.
To estimate the effect of this asymmetry, we first consider Equation~(\ref{eq:newalg}) in a mean field fashion, where each element of  $M_{cp}$ is set to the average value $\langle M \rangle = \sum_{c,p}M_{cp}/\mathcal{C}\mathcal{P}$, and write, at the fixed point:
\begin{equation}
  \displaystyle
  \left\{
    \begin{array}{ll}
      \tilde{f} = \delta^2+ \mathcal{P} \langle M \rangle\, \tilde{p}^{-1}\\
      \tilde{p} = 1 + \mathcal{C} \langle M \rangle\, \tilde{f}^{-1},
    \end{array}
  \right.
\label{eq:newalg2mf}
\end{equation}
with now all $\tilde{F}_c$ and $\tilde{P}_p$ set to be equal to their mean field value $\tilde{f}$ and $\tilde{p}$, respectively. By setting $\delta=0$, we find $\tilde{p} = 1/(1-\frac{\mathcal{C}}{\mathcal{P}})\approx 1+\frac{\mathcal{C}}{\mathcal{P}}$ and $\tilde{f}=\mathcal{P}-\mathcal{C}$.
Indeed, an approximate expression for the fixed point of Equation~(\ref{eq:newalg2}) in the regime $\delta \ll 1$ and $\mathcal{C} \ll \mathcal{P}$ can be derived also beyond the mean field approximation. To this end, we set again $\delta=0$ and consider the corresponding fixed point equation associated to Equation~(\ref{eq:newalg2}), i.e.:
\begin{equation}
	\displaystyle
	\left\{
	\begin{array}{ll}
		\tilde{F}_c = \sum_{p'} M_{cp'}/\tilde{P}_{p'} & \mbox{with}~~~ 1\le c \le \mathcal{C}\\
		\tilde{P}_p = 1+ \sum_{c'} M_{c'p}/\tilde{F}_{c'} & \mbox{with}~~~ 1\le p \le \mathcal{P}.\\
	\end{array}
	\right.
	\label{eq:newalg3}
\end{equation}
From the empirical structure of the matrix $\mathbf{M}$, we observe that the quantity $D_c = \sum_p M_{c,p}$, representing the diversification of country $c$, i.e., the number of different products exported by $c$, is of the order of $\mathcal{P}$, at least for the majority of countries (as an average over all the years considered we find that 70\% of the countries have $0.1 \le D_c / \mathcal{P} \le 1$).
Therefore, setting $\tilde P^* = \max_p \tilde P_p$ and $\tilde F_* = \min_c \tilde F_c$, Equation~(\ref{eq:newalg3}) implies:
\[
	\displaystyle
\left\{
\begin{array}{ll}
	\tilde{F}_c \ge D_c / \tilde P^* \approx \mathrm{const}\, \mathcal{P} / \tilde P^* & \mbox{with}~~~ 1\le c \le \mathcal{C}\\
	\tilde P^* \le 1 + \mathcal{C} / \tilde F_*. & \\
\end{array}
\right.
\]
From the first estimate, $\tilde F_* \ge \mathrm{const}\, \mathcal{P} / \tilde P^*$,  and therefore, by the second estimate, $\tilde P^* \le 1 + \mathrm{const}\,\frac{\mathcal{C}}{\mathcal{P}} \tilde P^*$. As $P_p \ge 1$, we conclude that $\tilde P_p=1+W_p$ with $W_p$ in the order of magnitude of $\mathcal{C}/\mathcal{P}$, and, as a consequence, $\tilde F_c$ is of the order of magnitude of $\mathcal{P}$.

We next compute explicitly the values of $\tilde F_c$ and $\tilde P_p$ at the first order in this approximation.
The calculation of second order terms can be found in Appendix~\ref{app:secondorder}.
By using the first order approximation $(1+a)^{-1}\approx 1-a$ twice, from Equation~(\ref{eq:newalg3}) we have:
\[
W_p \approx\; \sum_{c'} \frac{M_{c'p}}{D_{c'}} \left(1 + \frac{1}{D_{c'}} \sum_{p'} M_{c'p'} W_{p'}\right).
\]
Now let $\mathbf{H}$ be the square matrix of elements $H_{pp'} = \sum_{c'} M^T_{pc'} D_{c'}^{-2} M_{c'p'}$. Letting $D^{-1}$ be the column vector with components $1/D_c$ and $\mathbf{1}$ the identity matrix, the last displayed formula reads:
\[
 (\mathbf{1}  - \mathbf{H}) W \approx \mathbf{M}^T D^{-1}.
\]
We now observe that:
\(
H_{pp'} \le \sum_{c'}1/D_{c'}^2 \le \mathrm{const}\, \mathcal{C}/\mathcal{P}^2
\).
Therefore, the matrix $(\mathbf{1}-\mathbf{H})$ is close to the identity (the correction is of order $\mathcal{C}/\mathcal{P}^2$) and hence invertible (with also the inverse close to the identity). In this approximation, $W = \mathbf{M}^T D^{-1}$, so that the rescaled (reciprocals of the) qualities of products are given by:
\begin{equation}
\label{eq:Papprox}
\tilde P = 1 + \mathbf{M}^T D^{-1}.
\end{equation}
In the same approximation, we obtain the rescaled fitnesses $\tilde F_c$; since:
\[
\tilde F_c = \sum_{p'} \frac{M_{cp'}}{1+W_{p'}} \approx \sum_{p'} M_{cp'}(1-W_{p'}),
\]
we have:
\begin{equation}
\label{eq:Fapprox}
\tilde F = D -\mathbf{K} D^{-1},
\end{equation}
having introduced the \emph{co-production} matrix $\mathbf{K} = \mathbf{M} \mathbf{M}^T$ with elements $K_{cc'} = \sum_{p'} M_{cp'}M_{c'p'}$, representing the number of the same products exported by the two countries $c$ and $c'$.

\begin{figure}[t]
\centering
 \includegraphics[width=7.5cm]{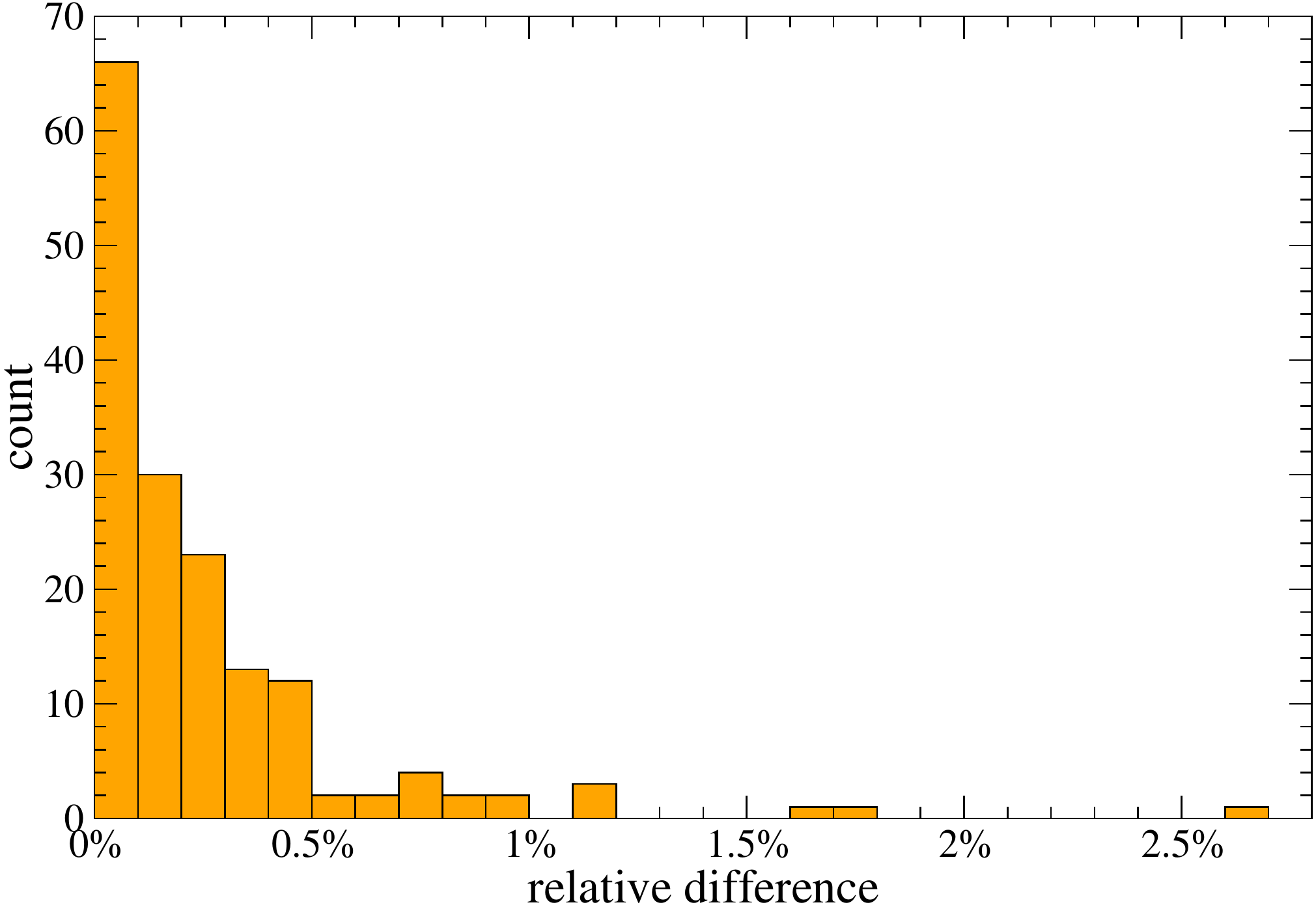}
\caption{ {Numerical vs Analytic relative error:}
  The histogram of the relative difference $(\tilde{F}_{c}^\mathrm{(fixed~point)} -\tilde{F}_{c}^\mathrm{(approximated)})/\tilde{F}_{c}^\mathrm{(fixed~point)}$ is plotted with the number of countries on the vertical axis. The approximated values are calculated using Equation~(\ref{eq:Fapprox}).
  \label{fig:num_vs_anal}
}
\end{figure}

It is interesting to note how, up to the first order approximation, the values of the fitness of countries are depending on the co-production matrix and diversification only.
The goodness of the approximations above can be appreciated in Figure \ref{fig:num_vs_anal} that shows how the relative difference between the numerical values at the fixed point and the approximate solution of Equation~(\ref{eq:Fapprox}) is below 0.5\% for more than 85\% of the countries.

It is worth noting that in a recent work the Economic Complexity Index (ECI) defined in Ref.~\cite{hidalgo2009} has been connected to the spectral properties of a weighted similarity matrix $\mathbf{\tilde{M}}$ resembling our co-production matrix $\mathbf{K}$ \cite{mealy2017}.
This similarity is only apparent since in ECI the matrix $\mathbf{\tilde{M}}$ is defined as:
\[
    \tilde{M}_{cc'} = \sum_p M_{cp}M_{c'p}/D_c U_p~~~\text{with}~~~U_p=\sum_c M_{cp}~\text{ubiquity of product}~p,
\]
i.e., it contains a further weighting term (the ubiquity) in the sum defining it.
Besides, the two metrics of ECI and Fitness-Complexity differ very much from each other: ECI relies on a linear homogeneous map, while Fitness-Complexity relies on a non-linear and in this work also non-homogeneous map.

\subsection{Country Inefficiency and 	Net-Efficiency}

\begin{figure}[t]
\centering
 \includegraphics[width=7.5cm]{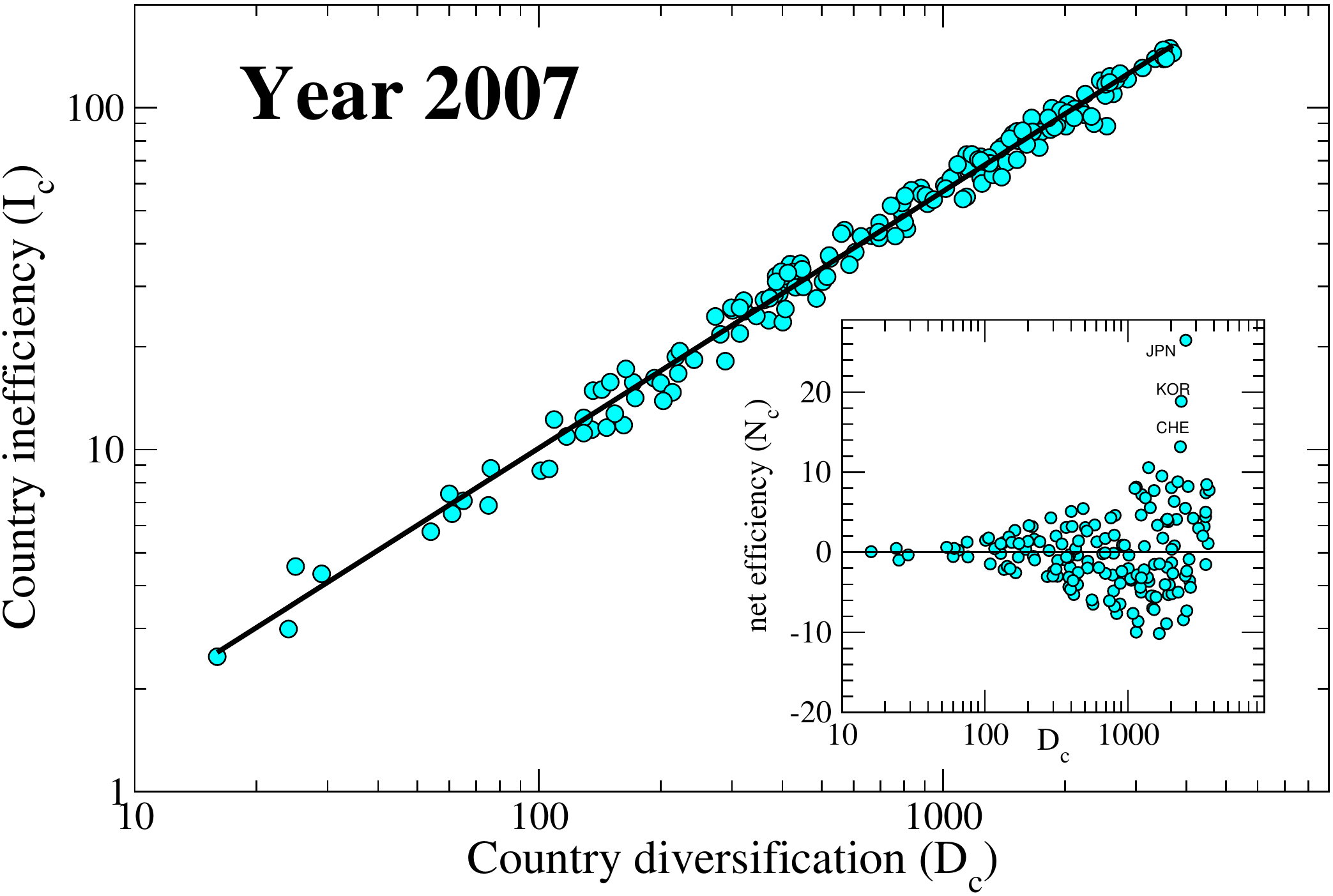}
\caption{ {Role of diversification:}
The country inefficiency $I_c=D_c-\tilde{F}_c$ is plotted vs. the diversification $D_c$ with the black line representing the power-law relation $I_c \approx D_c^{\,0.75}$ (linear regression with correlation coefficient 0.994).
In the inset the net efficiency $N_c$, defined as the difference between the black line and the inefficiency of the main graph, is shown.
Plotted data pertain to year 2007.
We find a similar behaviour for all the years considered with the exponent of $D_c$ between 0.73 and 0.76, and the correlation coefficient between 0.993 and 0.995.
  \label{fig:diversification}
}
\end{figure}

\begin{figure}[t]
\centering
 \includegraphics[width=8cm]{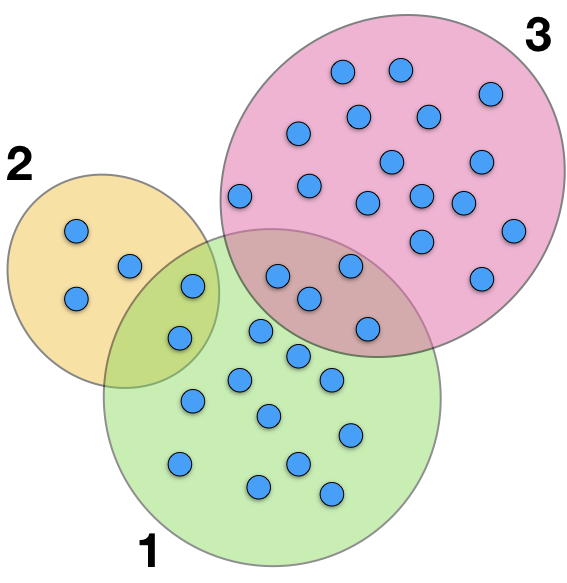}
\caption{ \textbf{Inefficiency cartoon:}
    Large ovals represent three countries, while small circles represent products.
    In this simple example, the inefficiency $I_1$ of country 1 is $I_1=K_{12}/D_2 + K_{13}/D_3$.
    From the figure we get $K_{12}=2$ and $K_{13}=4$, i.e.\ the number of products exported by both countries (the cardinality of the intersection sets), and the diversifications $D_1=17$, $D_2=5$, $D_3=20$.
    Thus, $I_1=2/5+4/20=0.6$ and the approximated fitness $\tilde{F}_1\approx16.4$.
  \label{fig:fitness-cartoon}
}
\end{figure}

From Equation~(\ref{eq:Fapprox}) we deduce that the leading part of fitness $\tilde{F}_c$ is given by the diversification $D_c$. The diversification of a country is indeed an important quantity, for the calculation of which we do not need any complicated algorithm.
On the other hand, what the non-linear map proposed does, is to quantify how a country manages to successfully differentiate its products, and indirectly offers an estimate of the capabilities of a nation.
In fact, a country exporting mainly raw materials would be less efficient with respect to a country exporting high technological goods, when they have the same diversification value.
For this reason, we introduce the new quantity $I_c = D_c-\tilde{F}_c$, inefficiency of country $c$: the smaller the value $I_c$ the more efficient is the diversification it chooses.
From the approximate solution displayed in Equation~(\ref{eq:Fapprox}), we get that $I_c\approx \sum_{c'} K_{cc'}/D_{c'}$, so that the inefficiency of a country is a weighted average of its co-production matrix elements.
The dependence of the country inefficiency on the diversification is displayed in {Figure~\ref{fig:diversification}}, while a visual representation of it is displayed in {Figure~\ref{fig:fitness-cartoon}}. 

It is interesting to notice how a clear power-law dependence exists between the inefficiency and the diversification of a country.

\begin{figure}[t]
\centering
 \includegraphics[width=0.49\textwidth]{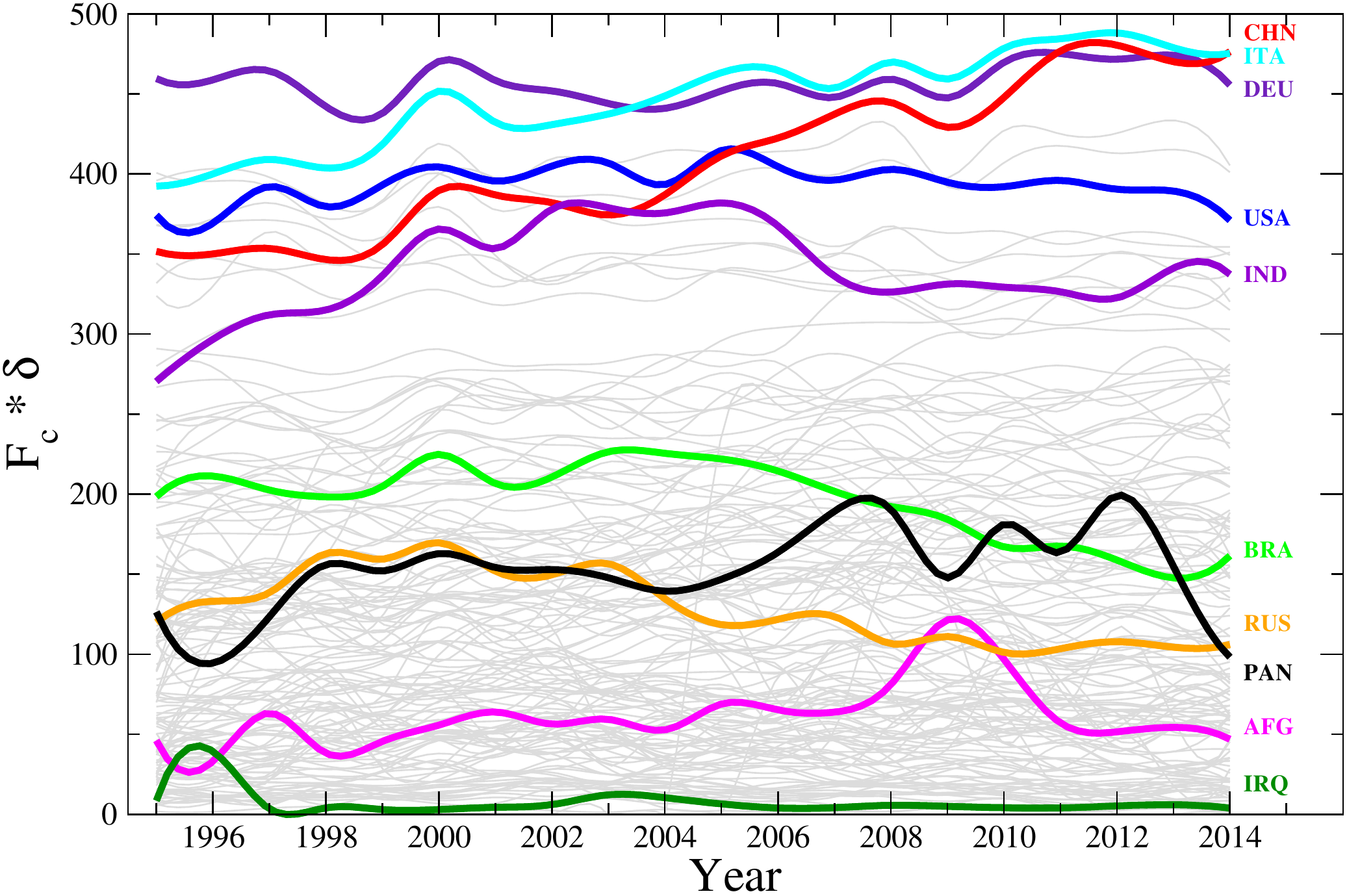}%
 \includegraphics[width=0.49\textwidth]{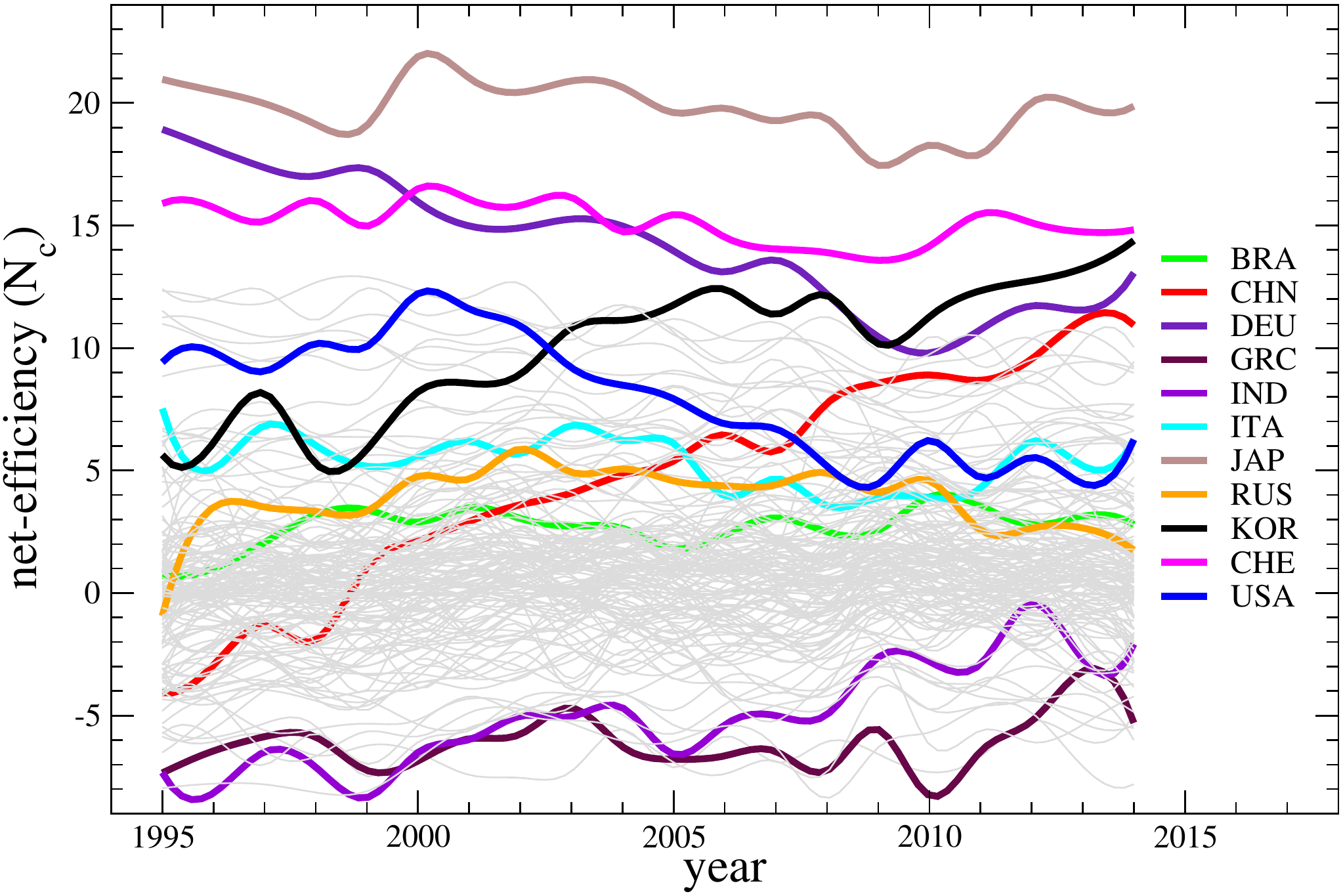}
\caption{{Time evolution of fitness and net-efficiency:}
  (Left panel) Country fitness yearly evolution as estimated by the new metric.
  (Right panel) yearly time evolution of country net efficiency. The net efficiency is a detrended version of the inefficiency defined in the text and displayed in the inset of Figure~\ref{fig:diversification} for the year 2007.
  Curves were artificially smoothed by a cubic spline for a better visual representation.
  \label{fig:timeevolution}
}
\end{figure}

The structure of the $\mathbf{M}$ matrix is such that those countries with high diversification also export low quality goods in average.
Therefore to a large diversification would statistically correspond a large inefficiency, though the found power-law is not trivial and depends on the structure of the $\mathbf{M}$.
A similar power-law behaviour is found between the fitness calculated with the traditional metric and the diversification, but with a different exponent (from the left panel of Figure~\ref{fig:comparison} we deduce that there is a power-law relation between the fitnesses calculated with the original metric and this new metric, and the exponent is around 1.53; since the fitness $F_c$ calculated with the new metric goes as $D_c$ at the first order, then the old fitnesses also go as $D_c^{\,1.53}$).
In order to better appreciate the production strategies of countries, we subtracted the common power-law trend of the dependency of the inefficiency on the diversification for each year, changed its sign and plotted the result in the right panel of Figure~\ref{fig:timeevolution}, which thus shows the time evolution of a quantity that we call country \emph{net-efficiency} $N_c$ (\emph{net} in the sense opposed to \emph{gross}) over the years 1995--2014.
It interesting to note how countries behave differently over the time lapse considered.
Some countries display a decreasing net-efficiency, others an increasing or a constant one.
What many of these curves have in common is the decreasing set up around year 2000, more pronounced in the case of higher developed countries, which lie at high net-efficiency in the graph.

\subsection{Local Convergence}

From the simulations it is clear that the fixed point obtained by iterating Equation~(\ref{eq:newalg}) is locally stable. We can also prove it by resorting to the Jacobian of the transformation, in the case of countries and products. First we recall that the sum over the indexes $c$ and $p$ of Equation~(\ref{eq:newalg}) run from $1$ to $\mathcal{C}$ and $\mathcal{P}$, respectively, with usually $\mathcal{C}\ll\mathcal{P}$.
In the case of countries and products $\mathcal{C}/\mathcal{P}\approx 10^{-1}$. We also fix $\phi_c=\pi_p=\delta \ll 1$, so that the fitnesses and the (reciprocals of the) qualities at the fixed point are approximately given by:
$F_c = \tilde F_c / \delta$ and $P_p = \delta \tilde P_p$ with $\tilde F_c$ and $\tilde P_p$ the components of the vectors $\tilde F$ and $\tilde P$ given in Equations~(\ref{eq:Fapprox}) and ~(\ref{eq:Papprox}) respectively.

Next, we calculate the Jacobian of the transformation at the fixed point, which can be simply expressed as the block anti-diagonal matrix:
\begin{equation}
 \mathbf{J} =
 \left( \begin{array}{cc}
         \mathbf{0} & -\mathbf{M}^T \mathbf{F}^{-2}\\
         -\mathbf{M} \mathbf{P}^{-2} & \mathbf{0}\\
        \end{array}
 \right),
\label{eq:jacob}
\end{equation}
having introduced the diagonal matrices $\mathbf{F}=\mathrm{diag}(F_1,F_2,\ldots,F_c)$ and $\mathbf{P}=\mathrm{diag}(P_1,P_2,\ldots,P_p)$, respectively.
We claim that the spectral radius $\rho(\mathbf{J})$ of the square matrix $\mathbf{J}$ is strictly smaller than one. Denoting by $\sigma(\mathbf{J})$ the spectrum of $\mathbf{J}$, this means that $\rho(\mathbf{J}) := \max\{|\lambda|\colon \lambda\in\sigma(\mathbf{J})\} < 1$.
From this it follows~\cite{hasselblatt} that the fixed point is asymptotically stable and the convergence exponentially fast. To prove the claim we consider the square of the Jacobian that can be written as a block diagonal matrix,
\begin{equation}
 \mathbf{J}^2 =
 \left( \begin{array}{cc}
         \mathbf{M}^T \mathbf{F}^{-2}\mathbf{M} \mathbf{P}^{-2} & \mathbf{0}\\
         \mathbf{0} & \mathbf{M} \mathbf{P}^{-2} \mathbf{M}^T \mathbf{F}^{-2}\\
        \end{array}
 \right),
\label{eq:jacobian2}
\end{equation}
and note that the traces of the two matrices on the diagonal is the same by applying a cyclic permutation. Noticing that $F_cP_p = \tilde F_c \tilde P_p$ and using the approximate solutions in Equations~(\ref{eq:Fapprox}) and (\ref{eq:Papprox}), we find with simple algebra that:
\begin{equation}
\mathrm{Tr} (\mathbf{J}^2) = 2\sum_{c,p} \frac{M_{c,p}^2}{F_c^2P_p^2} \approx  2\sum_{c,p} \frac{M_{c,p}^2}{D_c^2} = 2 \sum_c \frac{1}{D_c} \approx  \frac{\mathcal{C}}{\mathcal{P}} < 1.
\label{eq:trace}
\end{equation}
Moreover, we can write the two non trivial matrices composing $\mathbf{J}^2$ as:
\begin{equation}
         \mathbf{M}^T \mathbf{F}^{-2}\mathbf{M} \mathbf{P}^{-2} =
	    \mathbf{P}(\mathbf{P}^{-1} \mathbf{M}^T \mathbf{F}^{-1})(\mathbf{F}^{-1} \mathbf{M} \mathbf{P}^{-1})\mathbf{P}^{-1} =
	    \mathbf{P}\mathbf{A}^T\mathbf{A}\mathbf{P}^{-1}
\end{equation}
and:
\begin{equation}
         \mathbf{M} \mathbf{P}^{-2} \mathbf{M}^T \mathbf{F}^{-2} =
	\mathbf{F}(\mathbf{F}^{-1} \mathbf{M} \mathbf{P}^{-1})(\mathbf{P}^{-1} \mathbf{M}^T \mathbf{F}^{-1})\mathbf{F}^{-1} =
	\mathbf{F}\mathbf{A}\mathbf{A}^T\mathbf{F}^{-1},
\end{equation}
with $\mathbf{A} = \mathbf{F}^{-1} \mathbf{M} \mathbf{P}^{-1}$.
The matrices $\mathbf{A}\mathbf{A}^T$ and $\mathbf{A}^T\mathbf{A}$ are symmetric and positive-semidefinite so that their eigenvalues are real and non negative, and the matrices $\mathbf{F}\mathbf{A}\mathbf{A}^T\mathbf{F}^{-1}$ and $\mathbf{P}\mathbf{A}^T\mathbf{A}\mathbf{P}^{-1}$ have the same eigenvalues.
Therefore, the eigenvalues of $\mathbf{J}^2$ are real and non negative and we can write according to Equation~(\ref{eq:trace}):
\begin{equation}
 \mathrm{Tr}(\mathbf{J}^2) = \sum_i \lambda_i^2 < 1,
\end{equation}
with $\lambda_i$ eigenvalues of $\mathbf{J}$.
Finally, from the preceding equation we have $\max \lambda_i^2 < \max |\lambda_i| <1$ so that at the fixed point $\rho(\mathbf{J})<1$.

\subsection{Robustness to Noise}
\label{sec:noise}

\begin{figure}[t]
\centering
\includegraphics[width=8cm]{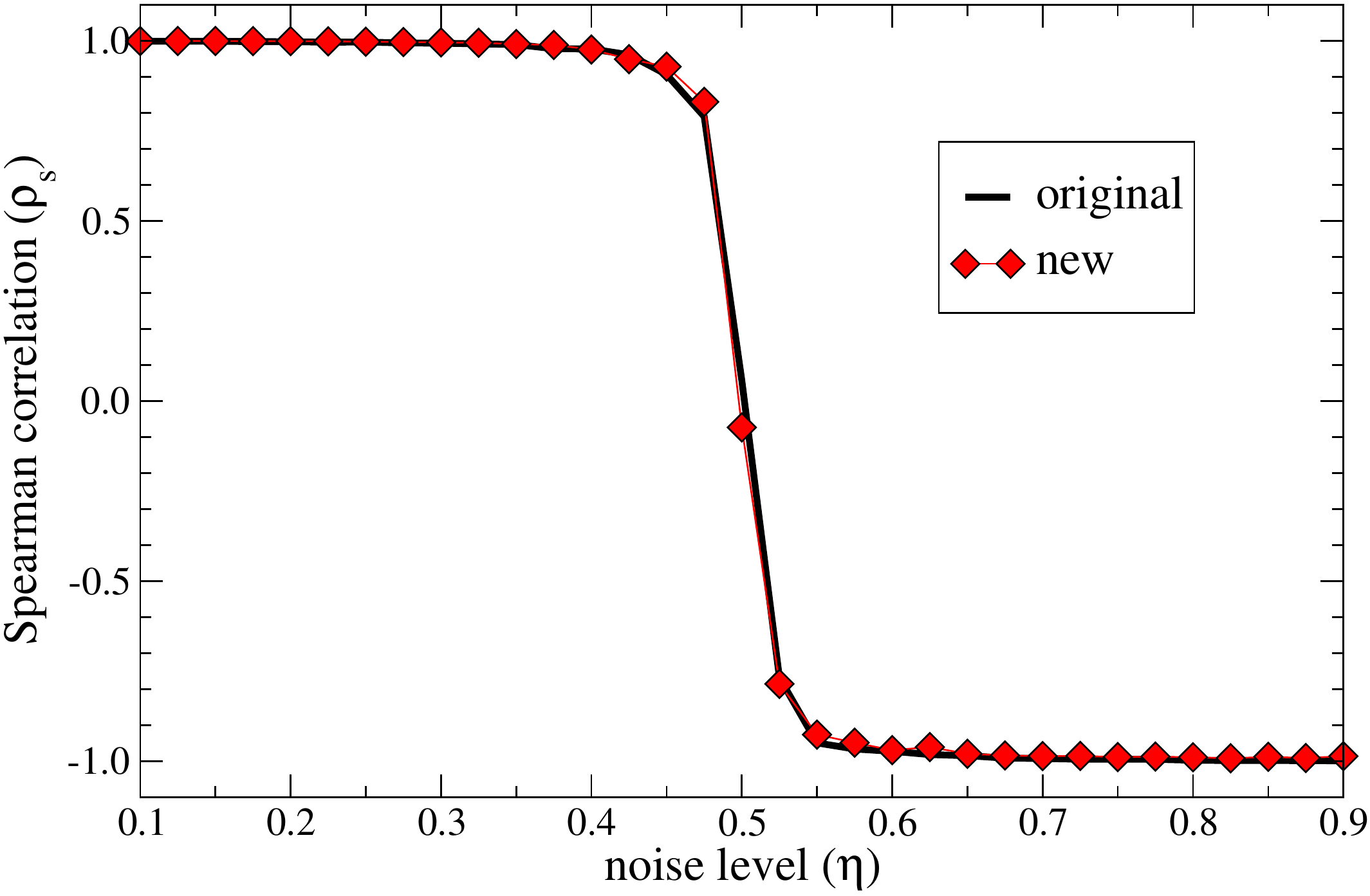}
\caption{ {Noise robustness:}
    Spearman correlation between the ranking of countries based on fitness at zero noise and at different noise levels $\eta$ (see Section~\ref{sec:noise} in the main text).
    The performance of the two metrics is practically indistinguishable.
    Note that at $\eta=1$ all the elements of matrix $\mathbf{M}$ are flipped so that the perturbed system is perfectly anti-correlated with the original one.
\label{fig:noise}}
\end{figure}

Fitness and complexity (quality) values depend on the structure of the matrix $M_{cp}$.
Noise can affect its elements by flipping their value.
Thus, we test the robustness of the new metric to noise as described in \cite{battiston2014metrics}.
The idea is to introduce random noise by flipping each single bit of the matrix with probability $\eta$, which then is a parameter tuning the noise level.
The rank of country fitnesses in presence of noise $R_c^\eta$ is then compared with the rank obtained without noise $R_c^0$.
The Spearman correlation $\rho_s$ is then evaluated between these two sets and shown in Figure~\ref{fig:noise} as a function of $\eta$ for both the original and the new metrics:
The new metrics show a perfect stability to random noise as the original ones with an unavoidable transition around $\eta\approx0.5$, where noise is so strong to alter significantly the structure of the matrix $M_{cp}$.

\section{Discussion}

The proposed new inhomogeneous non-linear metric to estimate economic fitness and complexity defined in Equations (\ref{eq:newalg}) and (\ref{eq:newalg2}) carries many advantages with respect to the original one.
The fitnesses and complexities resulting from these two approaches are not identical, but highly correlated to each other as witnessed by the plots in Figure~\ref{fig:comparison}.
This high correlation between the two metrics ensures that all the studies carried on with the original metric so far, can be obtained by applying this new metric as well.

Besides the stability of the metric and its robustness, one more advantage is that the fitness is well defined also for those countries that have low exportation volumes and that in the original metric had their fitness tending to zero.
For those countries it is now possible to undertake a comparative study based on hypothetical investments (changing the elements of the $\mathbf M$ matrix) so to make predictions on their economic impact.

By first symmetrising the original equations, by adding an inhomogeneous parameter and by rescaling the quantities, one obtains Equation (\ref{eq:newalg2}), where the parameter can be safely set to zero.
This ensures that this new metric is parameter free as the original one.
As a pleasant side effect, the fixed point of the map can be well approximated analytically, with an error with respect to the iterative fixed point of less than 3\% (see Figure~\ref{fig:num_vs_anal}).
The result is represented by Equations (\ref{eq:Papprox}) and (\ref{eq:Fapprox}) at the first order (Equations (\ref{eq:Papprox2}) and (\ref{eq:Fapprox2}) at the second order), which allow for a simple intuitive explanation of the complexity of products and fitness of countries.

Let us discuss Equation (\ref{eq:Fapprox}) first.
The result suggests that the fitness of a country is trivially related, at the first order, to its diversification: The more products a country exports, the larger is its fitness, i.e., the more developed its capabilities.
This simple explicit dependence of the fitness on the diversification is also an advantage with respect to the original metric, where the dependence was not explicitly clear.
The second term of Equation (\ref{eq:Fapprox}), which we call \emph{inefficiency}, is also very interesting.
If a country is the only one to export a given product, the contribution of this product to its fitness is a full one, or in other words, the contribution to the inefficiency is zero.
This situation mimics a condition of monopoly on that product and it is logical that the exporting country has the full benefit of it.
When a product is exported by multiple nations then it is critical to assess whether those countries export few or many other products (see Figure~\ref{fig:fitness-cartoon}).
If a product is exported by a country $c'$ with low diversification (low capabilities), then that product is not supposed to be of high complexity.
The result is that the ratio $K_{cc'}/D_{c'}$ can be close to one ($c=1, c'=2$ in the figure) and the inefficiency associated to the common products is high, resulting in a small contribution to the fitness of $c$.
The inefficiency can be interpreted in terms of the bipartite network of countries and products:
The $K_{cc'}$ counts the number of links that connect countries $c$ and $c'$ to the same products, while the differentiation $D_c$ is the node degree of country $c$.
In other words, for a country $c$ the inefficiency counts the links to common products of all other countries and weights them according to the degree of those.
To our knowledge, this kind of measure has never been considered in complex networks so far.
\begin{figure}[t]
\centering
 \includegraphics[width=10cm]{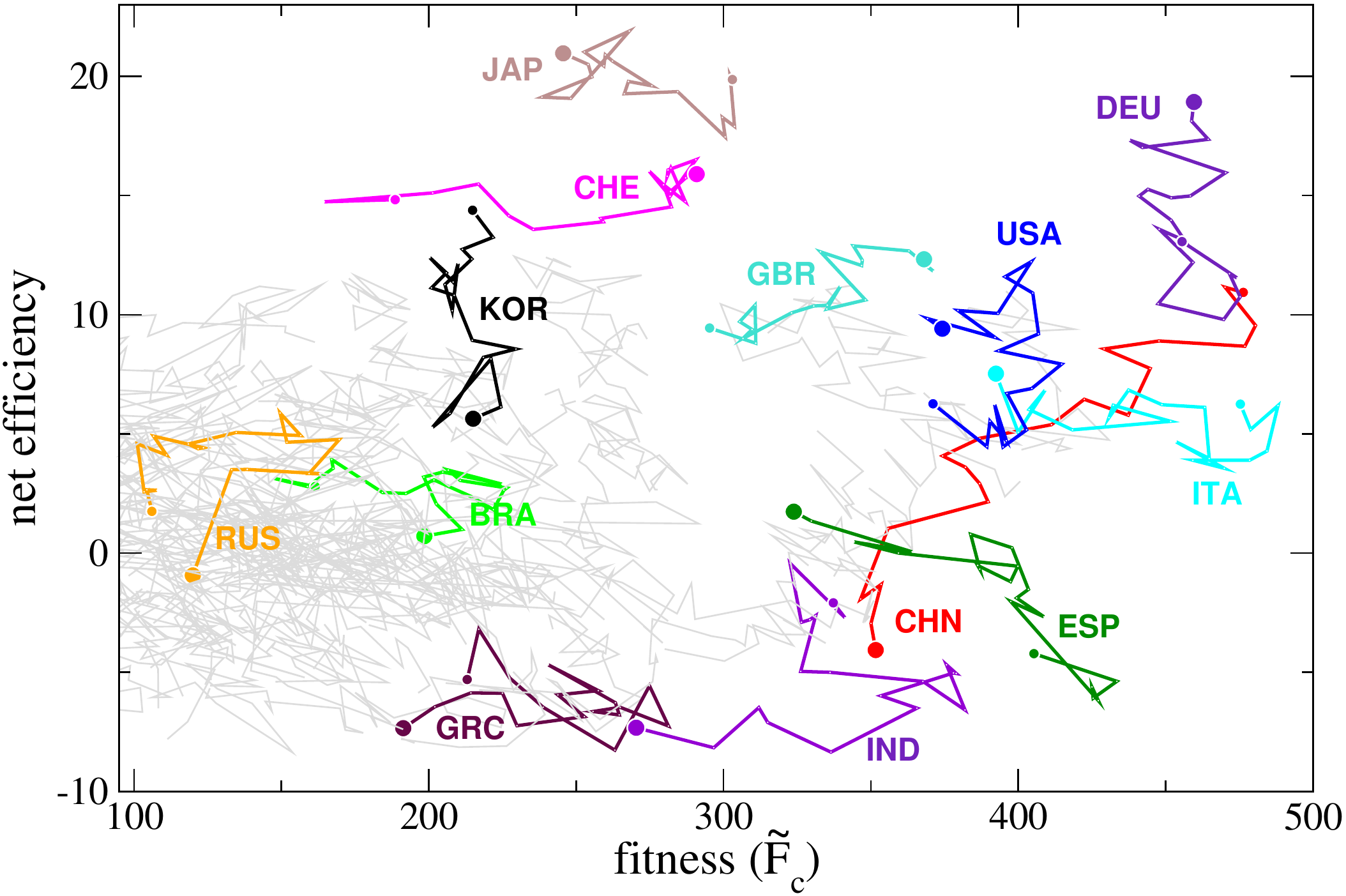}
\caption{ {Net-efficiency vs fitness:}
    each line corresponds to the time evolution of the connection between the fitness of a country and its net-efficiency in the period between 1995 and 2014.
    This figure connects the quantities on the vertical axes of the plots displayed in Figure~\ref{fig:timeevolution}.
    Lines start from a large circle (year 1995) and end with a small one (year 2014).
  \label{fig:neteff-fit}
}
\end{figure}
Since, statistically, countries with an high diversification also export many less complex products, the inefficiency is an increasing function of the diversification (Figure \ref{fig:diversification}, main graph).
If we subtract the general trend, which stems from the structure of the matrix $M_{cp}$, we can appreciate the net effect of selecting the goods to export.
We call this new de-trended quantity \emph{net-efficiency}.
In this way we somehow remove the negative effect of less valuable products and highlight the contribution of more sophisticated goods.
In the inset of Figure~\ref{fig:diversification} we show the net-efficiency as a function of diversification and underline the three nations (Japan, Korea, and Switzerland) that stand out among the others.
The time evolution of this new quantity is shown in the right panel of Figure~\ref{fig:timeevolution}.
We can combine the time evolution of both fitness and net-efficiency for a given country to determine to what degree they are correlated.
Figure~\ref{fig:neteff-fit} shows the two quantities for selected countries.
It is clear how these two quantities are not related to each other and represent two complementary information.
In fact, the fitness is mainly connected to the product diversification of a country (Equation~(\ref{eq:Fapprox})) while the net-efficiency is connected to how complex are the exported products.
In the figure we see two extreme cases represented by Switzerland (CHE) and South~Korea (KOR), whose lines are practically orthogonal.
While Switzerland have decreased the number of different exported goods in the years but have still exported complex products, South~Korea have kept the number of exported products as almost constant but have increased their complexity.
The opposite situation of South~Korea we notice for Germany (DEU), where the complexity of the exported goods has decreased in time.
Interestingly, China (CHN) has systematically increased both the number of exported goods and their complexity, which we interpret as a symptom of a solid economy in expansion.


The complexity of products is estimated by Equation (\ref{eq:Papprox}) as the reciprocal of the second term of the sum.
Since the diversification of a country $D_c$ is a direct measure of its capabilities, we expect to find a simple relation between it and the complexities of products $\tilde{Q}_p$.
Indeed, if we indicate with $c_i$ those countries exporting the product $p$, for which obviously we have $M_{c_i p}=1$, and with $m=\sum_c M_{cp}$, we can write:
\[
    \tilde{Q}_p \approx \left(
        \frac{1}{D_{c_1}} + \frac{1}{D_{c_2}} + \ldots + \frac{1}{D_{c_m}}
        \right)^{-1}
\]
from which we corroborate the main idea that the complexities of products are driven by the countries with low diversification (capabilities) that export it.
Just for amusement, we observe how the complexity of products can be considered as the equivalent resistor of a parallel of resistors each one with resistance $D_c$.
Somehow, a high $D_c$ represents an effective resistance to the creation of a product and its export, so that if a country exists with a low diversification exporting it, the effort (resistance) of producing that product is also low.


\section{Materials and Methods}
\subsection{Construction of the \textbf{M} Matrix}

We exploit the UN-COMTRADE data set~\cite{comtrade}, where re-export and re-import fluxes are explicitly declared, allowing us to exclude them from the analysis.
As reported by UNSTAT in Ref.~\cite{un2008}, the 81.8\% of the whole data set (96.8\% in case of developed countries) does not account for goods in transit.
Moreover, commodities that do not cross borders are not included in the data.

Given the export volumes $s_{cp}$ of a country $c$ in a product $p$ one can evaluate the Revealed Comparative Advantage (RCA) indicator \cite{RCA65} defined as the ratio:
\begin{equation}
 \mathrm{RCA}_{cp}=\frac{s_{cp}}{\sum_{c'} s_{c'p}} \left/ \frac{\sum_{p'} s_{cp'}}{\sum_{c'p'} s_{c'p'}}\right.
\end{equation}
in this way one can filter out size effects.
As described in the Supplementary information of \cite{tacchella2018dynamical}, from the time series of the RCA we can evaluate the productive competitiveness of each country in each product by assigning to it a productivity state from 1 to 4.
State 1 means that the country does not produce (or is very uncompetitive in producing) a product, state 4 means that it is one of the main producer in the world.
We can then project this states onto the binarized matrix $M_{cp}$ by simply setting its elements to unity whenever a state larger than 2 is encountered, and set them to null otherwise.

\section*{Acknowledgments}
{We acknowledge A.~Zaccaria and E.~Pugliese for interesting discussions.
We are grateful to an anonymous reviewer for suggesting to show Fig.~\ref{fig:neteff-fit}.
We acknowledge the EU FP7 Grant 611272 project GROWTHCOM and CNR PNR Project ``CRISIS Lab'' and
the Austrian Research Promotion Agency FFG under grant $\#$857136,  for financial support.
FFG is also covering the costs to publish in open access.}

\begin{appendices}
\section{Second order expansion of fitness and qualities}
\label{app:secondorder}
\noindent
In this section we compute explicitly the values of $\tilde F_c$ and $\tilde P_p$ for $\mathcal{C}\ll\mathcal{P}$ up to the second order of magnitude of $\mathcal{C}/\mathcal{P}$.
Letting $\varepsilon = \mathcal{C}/\mathcal{P}$, we expand $W_p = \varepsilon W_p^{(1)} + \varepsilon^2 W_p^{(2)} + O(\varepsilon^3)$. By assuming $D_c$ of the order of $\mathcal{P}$ and by using the second order approximation $(1+a)^{-1}\approx 1-a + a^2$ twice, Eq.~(\ref{eq:newalg3}) implies that
\begin{equation}
\label{eq:Fexp}
    \tilde{F}_c = D_c \bigg(1 - \varepsilon \sum_{p'} \frac{M_{cp'}}{D_{c}} W_{p'}^{(1)} - \varepsilon^2 \sum_{p'}  \frac{M_{cp'}}{D_{c}} \big[W_{p'}^{(2)} - (W_{p'}^{(1)})^2\big] + O(\varepsilon^3)\bigg), \qquad \mbox{with}~~~ 1\le c \le \mathcal{C},
\end{equation}
and
\[
\begin{split}
\varepsilon W_p^{(1)} + \varepsilon^2 W_p^{(2)} & = \sum_{c'} \frac{M_{c'p}}{D_{c'}} + \varepsilon \sum_{c',p'} \frac{M_{c'p}}{D_{c'}} \frac{ M_{c'p'}}{D_{c'}} W_{p'}^{(1)} + \varepsilon^2 \sum_{c',p'}\frac{M_{c'p}}{D_{c'}} \frac{ M_{c'p'}}{D_{c'}} \big[W_{p'}^{(2)} - (W_{p'}^{(1)})^2\big] \\ & \quad + \varepsilon^2 \sum_{c',p',p''} \frac{M_{c'p}}{D_{c'}} \frac{ M_{c'p'}}{D_{c'}}\frac{ M_{c'p''}}{D_{c'}} W_{p'}^{(1)}W_{p'}^{(1)} + O(\varepsilon^3), \qquad \mbox{with}~~~ 1\le p \le \mathcal{P}.
\end{split}
\]
By the assumption on the magnitude of $D_c$, the first sum in the right-hand side is of the order of $\varepsilon$, the second one is of the order of $\varepsilon^2$, while the last two sums are of the order $\varepsilon^3$. Therefore,
\[
\varepsilon W_p^{(1)} = \sum_{c'} \frac{M_{c'p}}{D_{c'}},\qquad \varepsilon^2 W_p^{(2)} = \sum_{c',p'} \frac{M_{c'p}}{D_{c'}} \frac{ M_{c'p'}}{D_{c'}} \varepsilon W_{p'}^{(1)}.
\]
Recalling $\mathbf{H}$ denotes the square matrix of elements $H_{pp'} = \sum_{c'} M^T_{pc'} D_{c'}^{-2} M_{c'p'}$ (hence $H_{pp'} \approx \varepsilon/\mathcal{P}$) and  $D^{-1}$ the column vector with components $1/D_c$, we have just showed that $W =  \mathbf{M}^T D^{-1} + \mathbf{H}\mathbf{M}^T D^{-1} + O(\varepsilon^3)$. Therefore, in the second order approximation, the rescaled (reciprocals of the) qualities of products are given by
\begin{equation}
	\label{eq:Papprox2}
	\tilde P = 1 + \mathbf{M}^T D^{-1} + \mathbf{H}\mathbf{M}^T D^{-1}.
\end{equation}
In the same approximation, from Eq.~(\ref{eq:Fexp}) we finally calculate the rescaled fitnesses $\tilde F_c$. Denoting by $(\mathbf{M}^T D^{-1})^2$ the column vector with components $(\mathbf{M}^T D^{-1})_p^2$ we get
\begin{equation}
	\label{eq:Fapprox2}
	\tilde F = D -\mathbf{K} D^{-1} + \mathbf{M} (\mathbf{M}^T D^{-1})^2 - \mathbf{M}\mathbf{H}\mathbf{M}^T D^{-1},
\end{equation}
where the \emph{co-production} matrix $\mathbf{K} = \mathbf{M} \mathbf{M}^T$ has been introduced just below Eq.~(\ref{eq:Fapprox}).

\section{Important quantities defined throughout the text}
\label{app:table}

\begin{tabularx}{\textwidth}{rX}
    $\mathcal{C}$, $\mathcal{P}$: &
    Total number of countries and products \\
    $\mathbf{M}$: &
    Binary matrix with element $M_{cp}=1$ if country $c$ is a competitive country in exporting product~$p$; $M_{cp}=0$ otherwise; export competitiveness is estimated by means of export volumes\\
    $F_c$, $Q_p$: &
    Fitness of country $c$ and quality (complexity) of product $p$ at the fixed point \\
    $P_p$: &
    Inverse of the quality of product $p$; it is a sort of product ``simplicity'' ($P_p=(Q_p)^{-1}$) \\
    $\delta$: &
    Inhomogeneous parameter; this parameter is crucial in achieving a stable algorithm to evaluate the metrics; it will be eventually let go to 0 to get a parameter free metric \\
    $\tilde{F}_c, \tilde{P}_p$: &
    Rescaled versions of the corresponding un-tilded quantities: $\tilde{F}_c=F_c\delta$, $\tilde{P}_p = P_p/\delta$; these quantities do not depend on $\delta$ as soon as $\delta\rightarrow0$ and are better suited to represent fitness and complexity rather than the un-tilded ones\\
    $\tilde{Q}_p$: &
    Similar to the complexity $Q_p$ above, but for the new metrics calculated with the inhomogeneous algorithm: $\tilde{Q}_p = (\tilde{P}_p -1 )^{-1}$\\
    $\mathbf{K}$: &
    Coproduction matrix with element $K_{cc'}$ equal to the number of the same products exported by countries $c$ and $c'$; $\mathbf{K} = \mathbf{M}\mathbf{M}^T$\\
    $D_c$: &
    Diversification of country $c$, i.e., the number of products the country $c$ is competitive in exporting\\
    %
    %
    $I_c$: &
    Inefficiency of country $c$ defined as $I_c=D_c-\tilde{F}_c$; it represents the fitness penalty resulting from exporting goods that are also exported by other countries\\
    $N_c$: &
    Net-efficiency of country $c$; it is a de-trended version of the inefficiency; in the dataset considered $N_c\approx D_c^{0.75}-I_c$; it represents how effectively a country diversifies its exported goods by focusing on products not exported by others, which are usually among the most complex \\
\end{tabularx}

\end{appendices}

\end{document}